\documentclass[11pt]{article}     

\begin{document}           % End of preamble and beginning of text.

\begin{center}

{\LARGE Tiled Image Convention for Storing Compressed Images}
{\Large in FITS Binary Tables}\\
\medskip
Richard L. White, STScI\\
Perry Greenfield, STScI\\
William Pence, NASA/GSFC\\
Doug Tody, NRAO\\
Rob Seaman, NOAO\\
\medskip
Version 2.2\\
May 4, 2011\\

\end{center}

\section{General Description}

This document describes a convention for compressing n-dimensional
images and storing the resulting byte stream in a variable-length
column in a FITS binary table.  The FITS file structure outlined
here is independent of the specific data compression algorithm that is
used.  The implementation details for 4 widely used compression 
algorithms are described here, but any other compression technique
could also be supported by this convention.

The general principle used in this convention is to first divide the
n-dimensional image into a rectangular grid of subimages or `tiles'.
Each tile is then compressed as a block of data, and the
resulting compressed byte stream is stored in a row of a variable
length column in a FITS binary table.  By dividing the image into tiles
it is generally possible to extract and uncompress subsections of the
image without having to uncompress the whole image.  The default
tiling pattern treats each row of a 2-dimensional image (or higher
dimensional cube) as a tile, such that each tile contains {\tt NAXIS1}
pixels.  This default many not be optimal for some applications or compression
algorithms, so any other rectangular tiling pattern may be defined using the
{\tt ZTILEn} keywords that are described below.  In the case of
relatively small images it may be sufficient to compress the entire
image as a single tile, resulting in an output binary table with 1
row.   In the case of 3-dimensional data cubes, it may be advantageous
to treat each plane of the cube as a separate tile if application
software typically needs to access the cube on a plane by plane basis.

\section{Keywords}

The following keywords are defined by this convention for use in the 
header of the FITS binary table extension to describe the structure
of the compressed image.

\begin{itemize}

\item {\tt ZIMAGE} (required keyword)  This keyword must have the
logical value T.  It indicates that the FITS binary table extension
contains a compressed image, and that logically this extension should
be interpreted as an image and not as a table.

\item {\tt ZCMPTYPE}  (required keyword) The value field of this
keyword shall contain a character string giving the name of the
algorithm that must be used to decompress the image.  Currently, values
of {\tt GZIP\_1}, {\tt RICE\_1}, {\tt PLIO\_1}, and {\tt HCOMPRESS\_1}
are reserved, and the corresponding algorithms are described in a later
section of this document.

\item {\tt ZBITPIX} (required keyword) The value field of this keyword
shall contain an integer that gives the value of the {\tt BITPIX} keyword in
the uncompressed FITS image.

\item {\tt ZNAXIS}  (required keyword) The value field of this keyword
shall contain an integer that gives the value of the {\tt NAXIS} keyword in
the uncompressed FITS image.

\item {\tt ZNAXISn} (required keywords)  The value field of these keywords
shall contain a positive integer that gives the value of the {\tt NAXISn}
keywords in the uncompressed FITS image.

\item {\tt ZTILEn} (optional keywords)  The value of these indexed
keywords (where {\tt n} ranges from 1 to {\tt ZNAXIS}) shall contain a
positive integer representing the number of pixels along axis {\tt n}
of the compression tiles. Each tile of pixels is compressed separately
and stored in a row of a
variable-length vector column in the binary table.  The size of each
image dimension (given by {\tt ZNAXISn}) is not required to be an integer
multiple of ZTILEn, and if it is not, then the last tile
along that dimension of the image will contain fewer image pixels than
the other tiles.  If the {\tt ZTILEn} keywords are not present then the
default 'row by row' tiling will be assumed such that {\tt ZTILE1 = ZNAXIS1},
and the value of all the other {\tt ZTILEn} keywords equals 1.

The compressed image tiles are stored in the binary table in the same
order that the first pixel in each tile appears in the FITS image; the
tile containing the first pixel in the image appears in the first row
of the table, and the tile containing the last pixel in the image
appears in the last row of the binary table.

\item {\tt ZNAMEn and ZVALn}  (optional keywords)  These pairs of
optional array keywords (where n is an integer index number starting
with 1)  supply the name and value, respectively, of any
algorithm-specific parameters that are needed to compress or uncompress
the image.  The value of {\tt ZVALn} may have any valid FITS datatype.
The order of the compression parameters may be significant, and may be
defined as part of the description of the specific decompression
algorithm.

\item {\tt ZMASKCMP} (optional keyword) Used to record the name of the
image compression algorithm that was used to compress the optional null pixel
data mask.  See the``Preserving undefined pixels with lossy compression''
section for more details.

\item  The following  8 optional keywords are defined to store a verbatim copy 
of the the value and comment fields of the corresponding keywords in
the original uncompressed FITS image.  These keywords can be used to reconstruct
an identical copy of the original FITS file when the image is uncompressed.

\begin{itemize}
\item {\tt ZSIMPLE}  - preserves the original {\tt SIMPLE} keyword 
\item {\tt ZTENSION} - preserves the original {\tt XTENSION} keyword
\item {\tt ZEXTEND} - preserves the original {\tt EXTEND} keyword
\item {\tt ZBLOCKED} - preserves the original {\tt BLOCKED} keyword
\item {\tt ZPCOUNT} - preserves the original {\tt PCOUNT} keyword 
\item {\tt ZGCOUNT} - preserves the original {\tt GCOUNT} keyword
\item {\tt ZHECKSUM} - preserves the original {\tt CHECKSUM} keyword
\item {\tt ZDATASUM} - preserves the original {\tt DATASUM} keyword

\end{itemize} 

The {\tt ZSIMPLE}, {\tt ZEXTEND}, and {\tt ZBLOCKED} keywords may only be
used if the original uncompressed image was contained in the primary array
of the FITS file.  The {\tt ZTENSION}, {\tt ZPCOUNT}, and {\tt ZGCOUNT}
keywords may only be used if the original uncompressed image was contained in 
in IMAGE extension.

\item {\tt ZQUANTIZ} (optional keyword) This keyword records the name of the
algorithm that was used to quantize floating-point image pixels into integer 
values which are then passed to the compression algorithm, as discussed further
in section 4 of this document.

\item{\tt Other Keywords}  The FITS header of the compressed image   may contain
other optional keywords.  If a FITS primary array or IMAGE extension is
compressed using the convention described here, it is recommended that 
all the keywords  in the
header of the original image, except for the mandatory keywords mentioned above,
be copied verbatim and in the same order into the header of the binary table 
extension that
contains the compressed image.  All these keywords will have the same meaning  and
interpretation as they did in the original image, even in cases where the keyword
is not normally expected to occur in the header of a
binary table extension  (e.g., the {\tt BSCALE} and {\tt BZERO} keywords,
or the World Coordinate System keywords such as {\tt CTYPEn, CRPIXn} and {\tt
CRVALn}).

\end{itemize}

\section{Columns}

The following columns in the FITS binary table are defined by this
convention. The order of the columns in the table is not significant.
The column names (given by the {\tt TTYPEn} keyword) are shown here in upper
case letters, but the case is not significant.

\begin{itemize}

\item {\tt COMPRESSED\_DATA}  (required column) Each row of this
variable-length column contains the byte stream that was generated as a
result of compressing the corresponding image tile.  The datatype of
the column (as given by the {\tt TFORMn} keyword) will generally be either
{\tt '1PB', '1PI'}, or {\tt '1PJ'}, depending on whether the compression algorithm
generates an output stream of 8-bit bytes, 16-bit integers, or 32-bit
integers, respectively.  If it is not possible to efficiently compress
a particular image tile, then the {\tt COMPRESSED\_DATA} vector in the
corresponding row will have a length of zero, and the uncompressed tile
pixels will be written instead to the {\tt UNCOMPRESSED\_DATA} or
{\tt GZIP\_COMPRESSED\_DATA} columns, as
described below.

\item {\tt UNCOMPRESSED\_DATA} (optional column)  This variable length
column contains the uncompressed pixels for any tiles that cannot
be compressed. The datatype of this column will usually correspond to
the datatype of the original image as shown in the following table:

\begin{center}
\begin{tabular}{|lrl|}
\hline
  Datatype   & {\tt BITPIX}    & {\tt TFORMn} \\
\hline
  byte          &     8            &          '1PB' \\
  short int     &    16            &          '1PI' \\
  long int      &    32            &          '1PJ' \\
  float         &   -32            &          '1PE' \\
  double        &   -64            &          '1PD' \\
\hline
\end{tabular}
\end{center}

If all the tiles in an image are able to be compressed, then the 
{\tt UNCOMPRESSED\_DATA} column is not required.  A tile compressed
image may only contain either
the {\tt UNCOMPRESSED\_DATA} column or the {\tt GZIP\_COMPRESSED\_DATA}
column (or neither), but not both.

\item {\tt GZIP\_COMPRESSED\_DATA} (optional column)  
The lossy quantization method that is often used to compress
floating-point images, as described in Section 4, can fail
in certain cases (for example, when all the pixels within a tile have the same
value and hence have a calculated RMS noise = 0).  In such cases,
the {\tt GZIP\_COMPRESSED\_DATA} column may be used to store
the original floating-point pixel values after compressing them with
the gzip algorithm.  This is almost always more efficient than storing
the uncompressed pixel values in the {\tt UNCOMPRESSED\_DATA} column.
This optional column was introduced in version 2.2 of this convention in 
May 2011.

If all the tiles in an image are able to be compressed, then the 
{\tt GZIP\_COMPRESSED\_DATA} column is not required.  A tile compressed
image may only contain either
the {\tt UNCOMPRESSED\_DATA} column or the {\tt GZIP\_COMPRESSED\_DATA}
column (or neither), but not both.

\item {\tt ZSCALE and ZZERO} (optional columns)  These columns give the
linear scale factor and zero point offset which may be needed to
transform the raw uncompressed values back to the original image pixel
values (or at least a close approximation to the original values) using
the following formula:

\begin{center}
       image\_pixel\_value = (uncompressed\_value * {\tt ZSCALE}) + {\tt ZZERO} \\
\end{center}

{\tt ZSCALE} and {\tt ZZERO} generally have double precision values and
have default values of 1.0 and 0.0, respectively. If the same values of
ZSCALE and ZZERO apply to every tile in the image, then they may be
given as header keywords rather than as table columns.

{\tt ZSCALE} and {\tt ZZERO} are typically used to scale floating-point
images (with {\tt BITPIX} = -32 or -64) into integers before
compression, since most compression algorithms are not very efficient
with floating-point data.  One particularly effective scaling algorithm 
is described in the next section.

These 2 parameters should not be confused with the reserved {\tt
BSCALE} and {\tt BZERO} keywords which may be present in integer FITS
images (which have BITPIX = 8, 16, or 32).   Any such integer images
should normally be compressed without any further scaling, and the {\tt
BSCALE} and {\tt BZERO} keywords should be copied verbatim into the
header of the binary table containing the compressed image.

\item {\tt ZBLANK} (optional column) In cases where floating-point
images are converted to integers before being compressed, this column
gives the the integer value that is used to represent undefined pixels
(if any) in the image.  These pixels would have an IEEE NaN (Not a
Number) value in the uncompressed floating-point FITS image.  If every
tile uses the same null value, then {\tt ZBLANK} may be given as a keyword
instead of as a table column.  If there are no undefined pixels in the
image then {\tt ZBLANK} is not required.  If the uncompressed image has an
integer datatype ({\tt ZBITPIX} $>$ 0) then the reserved {\tt BLANK} keyword which
already serves this purpose should be used instead of {\tt ZBLANK}.

\item {\tt NULL\_PIXEL\_MASK} (optional column)  In cases where the image 
contains undefined pixels and a lossy compression algorithm is used
(and hence the pixel values are not exactly preserved) then this column
is used to store a compressed image mask that records the location of
any undefined pixels.  See the ``Preserving undefined pixels with lossy compression''
section for more details.

\item {\tt Other Columns}  Any number of other columns may be present
in the table to supply other parameters that relate to each image tile.

\end{itemize}

\section{Quantization of Floating-Point Data}

Images that have floating-point data type pixels
often do not compress very effectively due to the presence of 
noise in the least significant bits of the pixel values.  In 
order to achieve a higher degree of compression, one can effectively
discard some of the noise bits by linearly scaling the image
into integer pixel values, so that

\begin{center}
       $F_i$ = ($I_i$ * {\tt ZSCALE}) + {\tt ZZERO} \\
\end{center}
\noindent
where $I_i$ and $F_i$ are the integer and floating-point values, respectively.

Note that the tiled image compression convention does not require that
floating point images be scaled to integers before compressing them, but if
linear scaling  is performed, then the ZSCALE and ZZERO columns in the FITS
binary table should be used to record the 2 scaling coefficients, as
described in the previous section.

The maximum amount of numerical precision will be preserved if the ZSCALE and
ZZERO values are calculated such that the scaled pixel values span the full
range of the integer datatype (e.g., from -32768 to +32767 for 16-bit
integers).  This may also preserve an undesirable amount of non-significant
noise, which can adversely affect the amount of compression that can be
achieved. 

A more effective scaling algorithm that preserves a specified amount of noise
in each pixel value is described by White and Greenfield (in the Proceedings
of the 1998 ADASS VIII conference) and by Pence, Seaman, and White, PASP 121,
414 (2009). With this method, the ZSCALE value (which is numerically equal to
the spacing between adjacent quantization levels) is calculated to be some
fraction, Q, of the RMS noise as measured in background regions of the
image.   It can be shown that the number of binary bits of noise that are
preserved in each pixel value is given by $log_2(Q) + 1.792$.  For example,
using Q = 8 (so that the quantized levels have a spacing of 1/8th of the
background RMS noise value) produces a quantized image that preserves about
4.8 bits of noise in each pixel. Specifying the quantization level relative
to the amount of noise in the image in this way produces comparable quality
images regardless of the noise level. Q is directly related to the
compressed  file size: decreasing Q by a factor of 2 will decrease the file
size by about 1 bit/pixel. In order to achieve the greatest amount of
compression, one should use the smallest value of Q that still preserves the
required amount of photometric and astrometric precision in the image.  

As the Q value is decreased, the spacing between the quantized levels in the
image increases, which can have the undesirable effect of significantly
biasing the  pixel values in the faint  regions the image (i.e., the 'sky'
level in typical astronomical images).   This bias can be mitigated by adding
noise during the quantization process.  So instead of simply scaling every
pixel value using the equation:
\begin{center}
   $I_i$ = ROUND(($F_i$ - {\tt ZZERO}) / {\tt ZSCALE}) \\
\end{center}

\noindent
(where the ROUND function rounds the result to the nearest integer value)
one can randomize the quantized levels by using this slightly modified equation:
\begin{center}
   $I_i$ = ROUND((($F_i$ - {\tt ZZERO}) / {\tt ZSCALE}) + $R_i$ - 0.5)\\
\end{center}

\noindent
where $R_i$ is a random number between 0 and 1, and the 0.5 is subtracted
so that the mean quantity is equal to 0.  Then when restoring the floating-point
value, the same random number is used with the inverse formula

\begin{center}
       $F_i$ = (($I_i$ - $R_i$ +0.5) * {\tt ZSCALE}) + {\tt ZZERO} \\
\end{center}

This technique, which is referred to as `subtractive dithering' in the signal
processing literature (e.g., "Quantization Noise" by Widrow and Kollar) has
the effect of dithering the zero-point of the quantization grid on a pixel by
pixel basis without introducing any additional noise in the image. The net
effect of this is that the mean (and median) pixel value in faint regions of
the image more closely approximate the value in the original unquantized
image than if all the pixels are scaled without randomization.  This can
significantly increase the precision when measuring the net flux from faint
sources in the compressed image.

The key requirement when using this technique is that the exact same random
number sequence must be used when quantizing the pixel values to integers,
and when restoring them to floating point values.  While most computer
languages supply a function for generating random numbers, these functions
are not guaranteed to generate the same sequence of numbers every time.
Accordingly, we define a specific algorithm here for generating a repeatable
sequence of pseudo random numbers.  The steps in the algorithm for quantizing
(or unquantizing) each tile of the image are as follows:

\begin{enumerate}
\item
Generate a sequence of 10000 random numbers using the algorithm given in 
Appendix A.  Since it would be computationally expensive to generate a  unique
random number for every pixel of large images, we  repeatedly recycle through
this `look up table'  of random numbers.

\item
The above sequence of random numbers is used when quantizing or unquantizing  each
tile of the floating point image. In order to avoid possible `banding' effects
if one were to use exactly the same sequence of  random numbers for every
tile, we calculate a unique, random offset to the first random number in the
sequence to use as a function of the tile number using  the formula:
\begin{center}
   offset = INT ( 500. * R(N) ) + 1 \\
\end{center}

\noindent
where offset is the ones-based index to the first random number in the sequence
to use, INT is the floating-point to integer truncation function, and R(N) is
the Nth random number in the sequence where N is the tile number.  If N exceeds
10000, then one should use ((N - 1) modulo 10000) + 1.  So for example, when
compressing the 2nd tile in an image, the 2nd random number  in the sequence
has a value of 0.131538, and thus the offset value is 66.  For reference,   the
66th random number should have a value of 0.493977.

This random number is then used to quantize (or unquantize) the first pixel of
this tile using the subtractive dithering function given above.  The next
random number in the sequence is then used for next pixel in the tile, and so
on.  

\item
If one reaches the end of the sequence of 10000 random numbers while quantizing
or unquantizing the pixels in  tile N, then one should cycle back through the
random  number sequence,  using a new random starting offset calculated using
the Nth + 1 random number.  For example, if one is quantizing tile number 9 of
the image, the original starting offset values would be calculated by
multiplying the 9th random number (0.679296) in the sequence by 500 (plus 1). 
Then if one reaches the end of the random number sequence again, the next
starting offset value is calculated using the 10th random number (0.934693). 
If necessary, this process is repeated using the next random number each time
(starting over at 1 if one reaches 10000).

\item
Repeat Steps 2 and 3 for each tile of the image.

\end{enumerate}

The above algorithm is clearly not unique, but we present it here as a
well defined method that should be easy to implement in almost any
computer language.  If this particular 'subtractive dithering' algorithm
is used when quantizing a floating point image, then the  following
keyword should be recorded in the compressed image header:

\begin{verbatim}
   ZQUANTIZ= 'SUBTRACTIVE_DITHER_1'
\end{verbatim}

Other values for this keyword may be defined in the future to
identify other quantizing methods.
If this keyword is not present in the header of a 
tile-compressed, quantized, floating-point
image, then it should be assumed that only simple linear scaling
was applied when quantizing the image.

It should be noted that an image that is quantized using
this technique can still be unquantized using the simple linear
scaling function.  The only side effect in this case is
to introduce slightly more noise in the image than if the full
subtractive dithering algorithm were applied.

\section{Preserving undefined pixels with lossy compression}

Any undefined pixels in a FITS image are flagged with a
special pixel value:  the BLANK keyword specifies the value 
in integer data type FITS images,
and an IEEE NaN (Not a Number) value is used in single or double precision floating
point FITS images. Floating point images are often converted to
scaled integers prior to compression (as described previously)
in which case the undefined pixel value is then given by the ZBLANK keyword 
(or column).

The  null pixel values in the image will be preserved
if a lossless compression algorithm is used.   If the image is 
compressed with a lossy
algorithm (e.g., H-Compress with a scale factor greater than 1), then
some other technique must be used to identify the null pixels in the image.

The recommended method of recording the null pixels when a lossy compression
algorithm is used is to create an integer data mask with the same 
dimensions as the image tile.  Set the null pixels to 1 and all the
other pixels to 0, then compress the mask array using a lossless 
algorithm such as PLIO or GZIP.  Store the compressed byte stream in a 
variable-length array 
column called 'NULL\_PIXEL\_MASK' in the row corresponding to that image tile.
The ZMASKCMP keyword should be used to record the name of the algorithm used 
to compress the data mask (e.g., RICE\_1).  
The data mask array pixels will be assumed to have
the shortest integer datatype that is supported by the compression algorithm
(i.e., usually 8-bit bytes).

When  uncompressing the image tile, the software must check if the
corresponding
compressed data mask exists with a length greater than 0, 
and if so, then uncompress the mask and
set the corresponding undefined pixels in the image array to the appropriate
value (as given by the BLANK or ZBLANK keyword).

\section{Currently Implemented Compression Algorithms}

This section describes the 4 compression algorithms that
are currently supported in the CFITSIO 
implementation  of this tiled image compression convention 
(available from the HEASARC web site).  This does
not imply that other implementations
of this convention must support these same
algorithms, nor does it limit other implementations from
supporting other compression algorithms.

\subsection{Rice compression algorithm}

The Rice algorithm (Rice, R. F., Yeh, P.-S., and Miller, W. H. 1993, 
in Proc. of the 9th AIAA Computing in Aerospace Conf., AIAA-93-4541-CP,
American Institute of Aeronautics and Astronautics)
is simple and very fast, 
compressing or decompressing $10^7$ pixels/sec on modern workstations. 
It requires only enough memory to hold a single block of 16 or 32 pixels at a time. 
It codes the pixels in small blocks and so is able to adapt very quickly to changes 
in the input image statistics (e.g., Rice has no problem handling cosmic rays, bright 
stars, saturated pixels, etc.).

The block size that is used should be recorded in the
compressed image header with

\begin{verbatim}
   ZNAMEn  = 'BLOCKSIZE'
   ZVALn   = 16 or 32
\end{verbatim}

\noindent
If these keywords are absent, then a default blocksize of 32 should be assumed.

The number of 8-bit bytes in each original integer pixel value should be recorded
in the compressed image header with

\begin{verbatim}
   ZNAMEn  = 'BYTEPIX'
   ZVALn   = 1, 2, 4, or 8
\end{verbatim}

\noindent
If these keywords are absent, then the default value of 4 bytes per pixel (32 bits)
should be assumed..

\subsection{GZIP compression algorithm}

Gzip is the compression algorithm used in the widely distributed GNU free software
utility of the same name.   It was created by Jean-loup Gailly and Mark
Adler. Version 0.1  was first publicly released on October 31, 1992. Version
1.0  followed in February 1993. It is based on the DEFLATE  algorithm, which
is a combination of LZ77 and Huffman coding.  DEFLATE was intended as a
replacement for LZW and other  patent-encumbered data compression algorithms
which, at the time,  limited the usability of compress and other popular
archivers.  Further information about this compression technique
is readily available on the Internet.

The gzip algorithm has no associated parameters
that need to be specified with the {\tt ZNAMEn and ZVALn} keywords.

\subsection{IRAF PLIO compression algorithm}

The IRAF PLIO (pixel list) algorithm was developed to store
integer-valued image masks in a compressed form.  Typical uses
of image masks are to segment images into regions, or to mark
bad pixels.  Such masks often have large regions of constant value
hence are highly compressible.  The compression algorithm used
is based on run-length encoding, with the ability to dynamically
follow level changes in the image, allowing a 16-bit encoding to
be used regardless of the image depth.   The  worst case
performance occurs when successive pixels have different values.
Even  in  this  case  the  encoding  will  only require  one  word
(16  bits) per mask pixel, provided either the delta intensity change
between pixels is usually less than  12  bits,  or  the mask
represents  a  zero floored step function of constant height.  The
worst case cannot exceed npix*2 words provided  the  mask  depth  is
24 bits or less.  

    A  good  compromise  between  storage  efficiency  and efficiency of
runtime access, while keeping things simple, is achieved if we  maintain
the  compressed  line  lists  as  variable  length  arrays of type short
integer (16 bits per list element), regardless of  the  mask  depth.   A
line  list  consists  of  a  series  of  simple  instructions  which are
executed in sequence to reconstruct a line of the  mask.   Each  16  bit
instruction  consists of the sign bit (not used at present), a three bit
opcode, and twelve bits of data, i.e.:
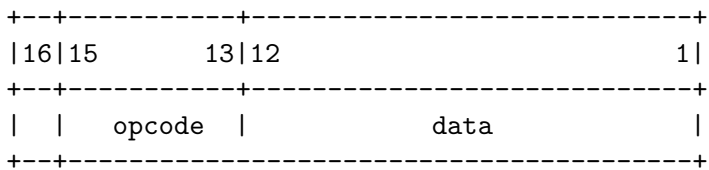
\begin{verbatim}
        +--+-----------+-----------------------------+
        |16|15       13|12                          1|
        +--+-----------+-----------------------------+
        |  |   opcode  |            data             |
        +--+-----------------------------------------+
\end{verbatim}
The  significance  of  the  data  depends  upon  the  instruction.   The 
instructions currently implemented are summarized in the table below.
\begin{verbatim}

     Instruction     Opcode           Description

        ZN            00        Output N zeros
        HN            04        Output N high values
        PN            05        Output N-1 zeros plus one high value
        SH            01        Set high value, absolute
        IH,DH         02,03     Increment or decrement high value
        IS,DS         06,07     Like IH-DH, plus output one high value
\end{verbatim}

In  order  to  reconstruct  a mask line, the application executing these
instructions is required to keep track of two values, the  current  high
value  and  the  current  position  in  the  output  line.  The detailed
operation of each instruction is as follows:
\begin{itemize}
\item[ZN]      Zero the next N (=data) output pixels.

\item[HN]      Set the next N output pixels to the current high value.

\item[PN]      Zero the next N-1 output pixels, and  set  pixel  N  to  the
            current high value.

\item[SH]      Set  the  high  value  (absolute  rather  than incremental),
            taking  the  high  15  bits  from  the  next  word  in   the 
            instruction  stream,  and  the  low 12 bits from the current
            data value.

\item[IH,DH]   Increment (IH) or decrement (DH) the current high  value  by
            the data value.  The current position is not affected.

\item[IS,DS]   Increment  (IS)  or decrement (DS) the current high value by
            the data value, and step, i.e., output one high value.

\end{itemize}

The high value is assumed to be set to 1 at the  beginning  of  a  line,
hence  the  IH,DH  and  IS,DS  instructions  are not normally needed for
Boolean masks.  If the length of a line segment  of  constant  value  or
the  difference  between  two  successive  high  values exceeds 4096 (12
bits), then multiple instructions are required to describe  the  segment
or intensity change.

\subsection{H-Compress algorithm}

Hcompress is an the image compression package
written by Richard L. White for use at the Space Telescope Science
Institute (rlw@stsci.edu).
Hcompress was used to compress the STScI Digitized Sky Survey and 
has also been used to compress the preview images in the Hubble Data Archive.
Briefly, the method used is:
\begin{enumerate}
\item
a wavelet transform called the H-transform (a Haar transform
		generalized to two dimensions), followed by
\item
quantization that discards noise in the image while retaining
		the signal on all scales, followed by
\item
quadtree coding of the quantized coefficients.
\end{enumerate}

The technique gives very good compression for astronomical images and
is relatively fast.  The calculations are carried
out using integer arithmetic and are entirely reversible.
Consequently, the program can be used for either lossy or lossless
compression, with no special approach needed for the lossless case
(e.g. there is no need for a file of residuals.) 

There are 2 user-defined parameters associated with the H-Compress algorithm:
an integer scale factor that determines the amount of compression, and a Boolean 
parameter the specifies whether the image should be smoothed during the
decompression operation, to reduce residual artifacts in the image.  

\begin{itemize}
\item  {\bf Scale Factor.} 
The integer scale parameter determines the  amount  of  compression.
Scale  =  0  or  1  leads  to lossless compression, i.e. the
decompressed image has exactly the same pixel values as  the
original image.  If the scale factor is greater than 1 then the compression is lossy:
the decompressed image will not be exactly the same  as  the
original.   For astronomical images, lossless compression is
generally rather ineffective because the images have a  good
deal of noise, which is inherently incompressible.  However,
if some of this noise is discarded then the images  compress
very  well.   The  scale  factor  determines how much of the
noise is discarded.  Setting scale  to  2
times  sigma, the RMS noise in the image, usually results in
compression by about a factor of  10  (i.e.  the  compressed
image  requires  about  1.5  bits/pixel),  while producing a
decompressed image that is nearly indistinguishable from the
original.    In   fact,   the  RMS  difference  between  the
decompressed image and the original image will be only about
1/2 sigma.  Experiments indicate that this level of loss
has no noticeable effect on either the visual appearance  of
the  image  or  on  quantitative analysis of the image (e.g.
measurements of positions and brightnesses of stars are  not
adversely affected.)

Using a larger value for scale results in higher compression
at the cost of larger differences between the compressed and
original images.  A rough rule of thumb  is  that  if  scale
equals  N  sigma,  then the image will compress to about 3/N
bits/pixel, and the RMS difference between the original  and
the  compressed  image  will be about N/4 sigma.  This crude
relationship is inaccurate both for  very  high  compression
ratios  and  for  lossless compression, but it does at least
give an indication of  what  to  expect  of  the  compressed
images.

For images in which the noise varies  from  pixel  to  pixel
(e.g.  CCD  images,  where  the noise is larger for brighter
pixels), the appropriate value for scale  is  determined  by
the  RMS  noise  level in the sky regions of the image.  For
images that are essentially noiseless, any lossy compression
is  noticeable  under  sufficiently  close inspection of the
image, but some loss is nonetheless acceptable  for  typical
applications.   Note  that  the  quantization scheme used in
Hcompress is not designed to give images that appear as much
like  the  original as possible to the human eye, but rather
is designed to produce images that are as similar as  possi-
ble  to the original under quantitative analysis.  Thus, the
emphasis is on discarding noise without affecting the signal
rather  than  on discarding components of the image that are
not very noticeable to the eye (as may be done, for example,
by  JPEG  compression.)  The resulting compression scheme is
not ideal for typical terrestrial images (though it is still
a  reasonably  good method for those images), but is believed
to be close to optimal for astronomical images.

It is not necessary to know what  scale factor was used
when compressing the image in order to uncompress it, but it is
still useful to record the value that was used.  It is recommended that 
the {\tt ZNAMEn} and {\tt ZVALn)} pair of keywords be used for this purpose,
with

\begin{verbatim}
    ZNAMEn = 'SCALE'
    ZVALn  =  I
\end{verbatim}

\noindent
where {\sl I} is the integer scale value.

\item
{\bf Smoothing Flag.}  
  At   high   compressions factors   the
decompressed image begins to appear blocky because
of the way information is discarded.  This blockiness
ness  is  greatly reduced, producing more pleasing
images,  if  the  image is smoothed slightly   during
decompression.  When done properly, the smoothing will
not affect any quantitative photometric or astrometric
measurements derived from the compressed image.  Of course,
the smoothing should never be applied when the image
has been losslessly compressed with a scale factor (defined above)
of 0 or 1.

The smoothing option only needs to be specified when uncompressing
the image, however, in many cases, this can best be determined
by the person or project that creates the compressed image files.
Thus it is recommended that the smoothing flag be specified
in the compressed image header with the {\tt ZNAMEn} and {\tt ZVALn} keywords
with

\begin{verbatim}
    ZNAMEn = 'SMOOTH'
    ZVALn  =  0 or 1
\end{verbatim}

\noindent
A value of 0 means no smoothing, and any other value means smoothing
is recommended.  This should be regarded as only a recommendation which
the image decompression program may override.

\end{itemize}

A paper describing Hcompress
was published in the Proceedings of the NASA Space and
Earth Science Data Compression Workshop, ed. James C. Tilton, Snowbird,
Utah, March 1992.  This paper is reproduced in the Appendix B of
this document.

\appendix
\section {Random Number Generator}

This portable random number generator algorithm comes from the publication 
``Random number generators: good ones are hard to find", Communications of the ACM,
Volume 31 ,  Issue 10  (October 1988) Pages: 1192 - 1201 which is available on the
Web.   This algorithm basically just repeatedly evaluates the function  seed = (a
* seed) mod m, where the values of a and m are shown below, but it is implemented
in a way to avoid integer overflow problems. 

\begin{verbatim}
  int random_generator(void) {

  /* initialize an array of random numbers */

    int ii;
    double a = 16807.0;
    double m = 2147483647.0;
    double temp, seed;
    float rand_value[10000];

    /* initialize the random numbers */
    seed = 1;
    for (ii = 0; ii < N_RANDOM; ii++) {
        temp = a * seed;
        seed = temp -m * ((int) (temp / m) );
        rand_value[ii] = seed / m;  /* divide by m to get value between 0 and 1 */
    }
  }
\end{verbatim}

If implemented correctly, the 10000th value of seed must equal 1043618065.

\newpage
\section {High-Performance Compression of Astronomical Images}
\bigskip
\centerline{\sl Richard L. White}
\smallskip
\centerline{\sl Joint Institute for Laboratory Astrophysics, University of Colorado}
\centerline{\sl Campus Box 440, Boulder, CO 80309}
\centerline{\sl and}
\centerline{\sl Space Telescope Science Institute}
\centerline{\sl 3700 San Martin Drive, Baltimore, MD 21218}
\smallskip
\centerline{\sl rlw@stsci.edu}
\bigskip
\centerline{\bf Summary}

Astronomical images have some rather unusual characteristics that make
many existing image compression techniques either ineffective or
inapplicable.  A typical image consists of a nearly flat background
sprinkled with point sources and occasional extended sources.  The
images are often noisy, so that lossless compression does not work very
well; furthermore, the images are usually subjected to stringent
quantitative analysis, so any lossy compression method must be proven
not to discard useful information, but must instead discard only the
noise.  Finally, the images can be extremely large.  For example, the
Space Telescope Science Institute has digitized photographic plates
covering the entire sky, generating 1500 images each having
$14000\times14000$ 16-bit pixels.  Several astronomical groups are now
constructing cameras with mosaics of large CCDs (each $2048\times2048$
or larger); these instruments will be used in projects that generate
data at a rate exceeding 100~MBytes every 5 minutes for many years.

An effective technique for image compression may be based on the H-transform/
(Fritze et al. 1977).  The method that we have developed can be used
for either lossless or lossy compression.  The digitized sky survey
images can be compressed by at least a factor of 10 with no noticeable
losses in the astrometric and photometric properties of the compressed
images.  The method has been designed to be computationally efficient:
compression or decompression of a $512\times512$ image requires only 4
seconds on a Sun SPARCstation~1.  The algorithm uses only integer
arithmetic, so it is completely reversible in its lossless mode, and it
could easily be implemented in hardware for space applications.

\smallskip
\centerline{\bf 1.  Introduction}

Astronomical images consist largely of empty sky.  Compression of such
images can reduce the volume of data that it is necessary to store (an
important consideration for large scale digital sky surveys) and can
shorten the time required to transmit images (useful for remote
observing or remote access to data archives.)

Data compression methods can be classified as either ``lossless''
(meaning that the original data can be reconstructed exactly from the
compressed data) or ``lossy'' (meaning that the uncompressed image is
not exactly the same as the original.)  Astronomers often insist that
they can accept only lossless compression, in part because of
conservatism, and in part because the familiar lossy compression
methods sacrifice some information that is needed for accurate
analysis of image data.  However, since all astronomical images contain
noise, which is inherently incompressible, lossy compression methods
produce much better compression results.

A simple example may make this clear.  One of the simplest data
compression techniques is run-length coding, in which runs of
consecutive pixels having the same value are compressed by storing the
pixel value and the repetition factor.  This method is used in the
standard compression scheme for facsimile transmissions.
Unfortunately, it is quite ineffective for lossless compression of
astronomical images because even though the sky is {\sl nearly}
constant, the noise in the sky ensures that only very short runs of
equal pixels occur.  The obvious way to make run-length coding more
effective is to force the sky to be exactly constant by setting all
pixels below a threshold (chosen to be just above the sky) to the mean
sky value.  However, then one has lost any information about objects
close to the detection limit.  One has also lost information about
local variations in the sky brightness, which severely limits the
accuracy of photometry and astrometry on faint objects.  Worse, there
may be extended, low surface brightness objects that are not detectable
in a single pixel but that are easily detected when the image is
smoothed over a number of pixels; such faint structures are
irretrievably lost when the image is thresholded to improve
compression.

\smallskip
\centerline{\bf 2.  The H-transform}

Fritze et al. (1977; see also Richter 1978 and Capaccioli et al. 1988)
have developed a much better compression method for astronomical images
based on what they call the {\sl H-transform} of the image.  A similar
transform called the S-transform has also been used for image
compression (Blume \& Fand 1989).  The H-transform is a two-dimensional
generalization of the Haar transform (Haar 1910).  The H-transform/ is
calculated for an image of size $2^N\times 2^N$ as follows:

\medskip
\begin{itemize}
\item Divide the image up into blocks of $2\times2$ pixels.  Call the
4 pixels in a block $a_{00}$, $a_{10}$, $a_{01}$, and $a_{11}$.

\item For each block compute 4 coefficients:

$
h_0 = (a_{11}+a_{10}+a_{01}+a_{00})/2 \\
h_x = (a_{11}+a_{10}-a_{01}-a_{00})/2 \\
h_y = (a_{11}-a_{10}+a_{01}-a_{00})/2 \\
h_c = (a_{11}-a_{10}-a_{01}+a_{00})/2 \\
$

\item
Construct a $2^{N-1}\times 2^{N-1}$ image from the $h_0$ values
for each $2\times2$ block.  Divide that image up into $2\times2$ blocks
and repeat the above calculation.  Repeat this process $N$ times,
reducing the image in size by a factor of 2 at each step, until only
one $h_0$ value remains.
\end{itemize}

\medskip
\noindent
This calculation can be easily inverted to recover the original image
from its transform.  The transform is exactly reversible using integer
arithmetic if one does not divide by 2 for the first set of
coefficients.  It is straightforward to extend the definition of the
transform so that it can be computed for non-square images that do not
have sides that are powers of 2.  The H-transform can be performed in place
in memory and is very fast to compute, requiring about $16M^2/3$
(integer) additions for a $M\times M$ image.

The H-transform is a simple 2-dimensional wavelet transform.  It has several
advantages over some other wavelet transforms that have been applied to
image compression (e.g., Daubechies 1988).  First, the transform can be
performed entirely with integer arithmetic, making it exactly
reversible.  Consequently it can be used for either lossless or lossy
compression (as indicated below) and one does not need a special
technique for the case of lossless compression (as was required, e.g.,
for the JPEG compression standard.)

A second major advantage is that the H-transform is a native 2-dimensional
wavelet transform.  The standard 1-dimensional wavelet transforms are
extended to two dimensions by transforming the image first along the
rows, then along the columns.  Unfortunately, this generates many
wavelet coefficients that are high frequency (hence localized) in the
$x$-direction but low frequency (hence global) in the $y$-direction.
Such coefficients are counter to the philosophy of the wavelet
transform:  high-frequency basis functions should be confined to a
relatively small area of the image.  Discarding these mixed-scale
terms, which may be negligible compared to the noise, generates very
objectionable artifacts around point sources and edges in the image.
The H-transform, on the other hand, is a fully 2-dimensional wavelet
transform, with all high frequency terms being completely localized.
It is consequently more suitable for image compression and produces
fewer artifacts.

A possible disadvantage of the H-transform is that other wavelet transforms
take better advantage of the continuity of pixel values within images,
so that they can produce higher compressions for very smooth images.
However, for astronomical images (which are mostly flat sky sprinkled
with point sources) the smoothness built into higher-order transforms
can actually reduce the effectiveness of compression, because one must
keep more coefficients to describe each point source.

\smallskip
\centerline{\bf 3. Compression Using the H-transform}

If the image is nearly noiseless, the H-transform is somewhat easier to
compress than the original image because the differences of adjacent
pixels (as computed in the H-transform) tend to be smaller than the original
pixel values for smooth images.  Consequently fewer bits are required
to store the values of the H-transform coefficients than are required for the
original image.  For very smooth images the pixel values may be
constant over large regions, leading to transform coefficients that
are zero over large areas.

Noisy images still do not compress well when transformed, though.
Suppose there is noise $\sigma$ in each pixel of the original image.
Then from simple propagation of errors, the noise in each of the H-transform
coefficients is also $\sigma$.  To compress noisy images, divide each
coefficient by $S\sigma$, where $S\sim1$ is chosen according to how
much loss is acceptable.  This reduces the noise in the transform to
$0.5/S$, so that large portions of the transform are zero (or nearly
zero) and the transform is highly compressible.

Why is this better than simply thresholding the original image?  As
discussed above, if we simply divide the image by $\sigma$ then we lose
all information on objects that are within $1\sigma$ of sky in a {\sl
single} pixel, but that are detectable by averaging a {\sl block} of
pixels.  On the other hand, in dividing the H-transform by $\sigma$, we
preserve the information on any object that is detectable by summing a
block of pixels!  The quantized H-transform preserves the mean of the image
for every block of pixels having a mean significantly different than
that of neighboring blocks of pixels.

As an example, Figure 1 shows a $128\times128$ section
($3.6\times3.6$~arcmin) from a digitized version of the Palomar
Observatory--National Geographic Society Sky Survey plate
containing the Coma cluster of galaxies.
Figures 2, 3, and 4 show the resulting
image for $S \simeq 0.5$, 1, and 2.  These images are compressed by
factors of 10, 20, and 60 using the coding scheme described below.  
In all cases a logarithmic gray scale is used to show the maximum detail
in the image near the sky background level; the noise is clearly
visible in Figure~1.  The
image compressed by a factor of 10 is hardly distinguishable from the
original.  In quantizing the H-transform we have adaptively filtered the
original image by discarding information on some scales and keeping
information on other scales.  This adaptive filtering is most apparent
for high compression factors (Fig.~4), where the sky has been smoothed
over large areas while the images of stars have hardly been affected.

%\topinsert
%\tabskip=0pt plus1fil
%\halign to\hsize{&\hbox to3.2truein{\hss #\hss}\cr
%\psfig{figure=orig.ps,width=3.8truein}&
%\psfig{figure=x10.ps,width=3.8truein}\cr
%{\smallsl Figure 1.  Original image of Coma cluster.}&
%{\smallsl Figure 2.  Result of compression by factor of 10.}\cr
%\noalign{\medskip}
%\psfig{figure=x20.ps,width=3.8truein}&
%\psfig{figure=x50.ps,width=3.8truein}\cr
%{\smallsl Figure 3.  Result of compression by factor of 20.}&
%{\smallsl Figure 4.  Result of compression by factor of 60.}\cr
%}
%\endinsert

The adaptive filtering is, in itself, of considerable interest as an
analytical tool for images (Capaccioli et al. 1988).  For example, one
can use the adaptive smoothing of the H-transform to smooth the sky without
affecting objects detected above the (locally determined) sky; then an
accurate sky value can be determined by reference to any nearby pixel.

The blockiness that is visible in Figure~4 is the result of difference
coefficients being set to zero over large areas, so that blocks of
pixels are replaced by their averages.  It is possible to eliminate the
blocks by an appropriate filtering of the image.  A simple but
effective filter can be derived by simply adjusting the H-transform
coefficients as the transform is inverted to produce a smooth image;
as long as changes in the coefficients are limited to $\pm S\sigma/2$,
the resulting image will still be consistent with the thresholded
H-transform.

\smallskip
\centerline{\bf 4. Efficient Coding}

The quantized H-transform has a rather peculiar structure.  Not only are large
areas of the transform image zero, but the non-zero values are strongly
concentrated in the lower-order coefficients.  The best approach we
have found to code the coefficient values efficiently is quadtree
coding of each bitplane of the transform array.  Quadtree coding has
been used for many purposes (see Samet 1984 for a review); the
particular form we are using was suggested by Huang and Bijaoui (1991)
for image compression.

\medskip
\begin{itemize}
\item
Divide the bitplane up into 4 quadrants.  For each quadrant code
a `1' if there are any 1-bits in the quadrant, else code a `0'.

\item Subdivide each quadrant that is not all zero into 4 more pieces
and code them similarly.  Continue until one is down to the level of
individual pixels.
\end{itemize}

\medskip
\noindent
This coding (which Huang and Bijauoi call ``hierarchic 4-bit one''
coding) is obviously very well suited to the H-transform image because
successively lower orders of the H-transform coefficients are located in
successively divided quadrants of the image.

We follow the quadtree coding with a fixed Huffman coding that uses 3
bits for quadtree values that are common (e.g., 0001, 0010, 0100, and
1000) and uses 4 or 5 bits for less common values.  This reduces the
final compressed file size by about 10\% at little computational cost.
Slightly better compression can be achieved by following quadtree
coding with arithmetic coding (Witten, Bell, and Cleary 1987), but the
CPU costs of arithmetic coding are not, in our view, justified for
3--4\% better compression.  We have also tried using arithmetic coding
directly on the H-transform,
with various contexts of neighboring pixels, but find it to be
both computationally inefficient and not significantly better than
quadtree coding.

For completely random bitplanes, quadtree coding can actually use more
storage than simply writing the bitplane directly; in that case we
just dump the bitplane with no coding.

Note that by coding the transform one bitplane at a time, the compressed
data can be viewed as an incremental description of the image.  One can
initially transmit a crude representation of the image using only the
small amount of data that is required for the sparsely populated, most
significant bit planes.  Then the lower bit planes can be added one by
one until the desired accuracy is required.  This could be useful, for
example, if the data is to be retrieved from a remote database --- one
could examine the crude version of the image (retrieved very quickly)
and abort the transmission of the rest of the data if the image is
judged to be uninteresting.

\smallskip
\centerline{\bf 5. Astrometric and Photometric Properties of Compressed Images}

Astronomical images are not simply subjected to visual examination,
but are also subjected to careful quantitative analysis.  For example,
for the image in Figure~1 one would typically like to do astrometric
(positional) measurements of objects to an accuracy much better than
1 pixel, photometric (brightness) measurements of objects to an accuracy
limited only by the detector response and the noise, and accurate
measurements of the surface brightness of extended sources.

We have done some experiments to study the degradation of astrometry and
photometry on the compressed images compared to the original images (White,
Postman, and Lattanzi 1991).  Even the most highly compressed images have very
good photometric properties for both point sources and extended sources;
indeed, photometry of extended objects can be improved by the adaptive
filtering of the H-transform (Capaccioli et al. 1988).  Astrometry is hardly
affected by the compression for modest compression factors (up to about a
factor of 20 for our digitized photographic plates), but does begin to degrade for
the most highly compressed images.

These results are based on tests carried out with tools optimized for
the original images; it is likely the best results will be obtained for
highly compressed images only with analysis tools specifically adapted
to the peculiar noise characteristics of the compressed images.

\smallskip
\centerline{\bf 6. Conclusions}

In order to construct the Guide Star Catalog for use in pointing the
Hubble Space Telescope, the Space Telescope Science Institute scanned
and digitized wide-field photographic plates covering the entire sky.
The digitized plates are of great utility, but to date it has been
impossible to distribute the scans because of the massive volume of
data involved (a total of about 600~Gbytes).  Using the compression
techniques described in this paper, we plan to distribute our digital
sky survey on CD-ROMs;  about 100 CD-ROMs will be required if the
survey is compressed by a factor of 10.

The algorithm described in this paper has been shown to be capable of
producing highly compressed images that are very faithful to the
original.  Algorithms designed to work on the original images can give
comparable results on object detection, astrometry, and photometry when
applied to the images compressed by a factor of 10 or possibly more.
Further experiments will determine more precisely just what errors are
introduced in the compressed data; it is possible that certain kinds of
analysis will give more accurate results on the compressed data than on
the original because of the adaptive filtering of the H-transform (Capaccioli
et al. 1988).

This compression algorithm can be applied to any image, not just to
digitized photographic plates.  Experiments on CCD images indicate that
lossless compression factors of 3--30 can be achieved depending on the
CCD characteristics (e.g., the readout noise).  A slightly modified
algorithm customized to the noise characteristics of the CCD will do
better.  This application will be explored in detail in the future.

We gratefully acknowledge grant from NAGW-2166 from the Science
Operations Branch of NASA headquarters which supported this work.
The Space Telescope Science Institute is operated by AURA with funding
from NASA and ESA.

\smallskip
\centerline{\bf References}

{\parskip=0pt \frenchspacing \advance\leftskip by\parindent \parindent=-\parindent

Blume, H., and Fand, A.  1989, {\sl SPIE Vol. 1091,
Medical Imaging III: Image Capture and Display}, p. 2.

Capaccioli, M., Held, E. V., Lorenz, H., Richter, G. M., and Ziener, R.
1988, {\sl Astronomische Nachrichten}, {\bf 309}, 69.

Daubechies, I. 1988, {\sl Comm. Pure and Appl. Math.}, {\bf 41}, 909.

Fritze, K., Lange, M., M\"ostl, G., Oleak, H., and Richter, G. M. 1977,
{\sl Astronomische Nachrichten},
{\bf 298}, 189.

Haar, A. 1910, {\sl Math. Ann.} {\bf 69}, 331.

Huang, L, and Bijaoui, A.  1991, {\sl Experimental Astronomy}, {\bf 1}, 311.

Richter, G. M. 1978,
{\sl Astronomische Nachrichten},
{\bf 299}, 283.

Samet, H. 1984, {\sl ACM Computing Surveys}, {\bf 16}, 187.

White, R. L., Postman, M., and Lattanzi, M. 1991, in {\sl Proceedings of the
   Edinburgh Meeting on Digital Sky Surveys}, in press.

Witten, I. H., Radford, M. N., and Cleary, J. G. 1987, {\sl Communications of
the ACM}, {\bf 30}, 520.

}

\end{document}